\documentclass{PoS}
\usepackage{graphicx}
\title{Volume dependence of Fisher's zeros }

\ShortTitle{Volume dependence of Fisher's zeros }

\author{\speaker{A. Denbleyker}, Daping Du, \speaker{Yuzhi Liu}, and {Y. Meurice}\\
Department of Physics and Astronomy, The University of Iowa,
Iowa City, Iowa 52242, USA\\
E-mail: \email{alan-denbleyker@uiowa.edu}\\
E-mail: \email{daping-du@uiowa.edu}\\
E-mail: \email{yuzhi-liu@uiowa.edu}\\
E-mail:\email{yannick-meurice@uiowa.edu}}
\author{A. Velytsky\\
EFI, University of Chicago, 5640 S. Ellis Ave., Chicago, IL 60637
and Argonne National Laboratory, 9700 Cass Ave., Argonne, IL 60439\\
E-mail:\email{vel@uchicago.edu} }

\abstract{We study the location of the partition function zeros in the complex $\beta$ plane (Fisher's Zeros) for SU(2) lattice gauge theory on $L^4$ lattices. 
We discuss recent attempts to locate complex zeros for $L=4$ and 6. 
We compare results obtained using 
various polynomial approximations of the logarithm of the density of states  and  a straightforward MC reweighting. We conclude that the method based on a combination of 
discrete Chebyshev orthogonality and patching plaquette distributions at different 
$\beta$ provides the more reliable estimates.}

\FullConference{The XXVI International Symposium on Lattice Field Theory \\
                July 14 - 19, 2008\\
                Williamsburg, Virginia, USA}

\begin{document}

\section{Introduction}
Locating the zeros of the partition function of lattice gauge theories in the complex $\beta$ plane and their volume dependence is important to understand the large order behavior of the weak coupling expansion \cite{ third,npp,quasi,lat07} at zero temperature and the nature of the finite temperature transition \cite{alves91,janke04}. These zeros are called Fisher's zeros 
\cite{fisher65} and should not be confused with 
Lee-Yang zeros which are zeros in the complex fugacity plane or the complex $\exp(-2\beta H )$ plane \cite{Lee-Yang} . 

In the following, we discuss the Fisher's zeros  of a pure gauge  $SU(2)$ theory 
with a partition function 
\begin{equation}
Z=\prod_{l}\int dU_l {\rm e}^{-\beta S} \  ,
\end{equation}
where S is the Wilson action
\begin{equation}
S=\sum_{p}(1-(1/N)Re Tr(U_p))\ , 
\end{equation} 
and $\beta=2N/g^2$. 

Our expectations is that at zero temperature, there is no singularity on the real axis 
of the complex $\beta$ plane and as the volume increases, the zeros stay away from 
the real axis. On the other hand at non-zero temperature, we expect that as the volume 
increases, the zeros pinch the real axis as for the 2D Ising model  \cite{fisher65}.

The zeros of the partition function can be calculated using the reweighting method \cite{falcioni81,alves91}.
\begin{equation}
Z(\beta_0+\Delta \beta)=Z(\beta_0)<\exp (-\Delta \beta S)>_{\beta_0}\ .
	\label{eq:pf}
\end{equation}
It is convenient to subtract $<S>_{\beta_0}$ from $S$ in the exponential because it removes fast 
oscillations without changing the complex zeros. 

$Z(\beta)$ is the Laplace transform of the density of states $n(S)$:
\begin{equation}
Z(\beta)=\int _0^{S_{max}}dS\ n(S)\exp(-\beta S)
\end{equation}
One can show \cite{gluodyn} that for $SU(2)$ $S_{max}=2\mathcal{N}_p$ and that for lattices with an even number of sites in each direction, $n(S)= n(2\mathcal{N}_p- S)$, where $\mathcal{N}_p$ is the number of plaquettes. For a  
$D$ dimensional cubic lattice with periodic boundary conditions,
$\mathcal{N}_p\equiv\ L^D D(D-1)/2$. When $n(S)$ is known, it is possible to calculate 
the partition function for any complex value of $\beta$. 

\begin{figure}
\includegraphics[width=2.in,angle=270]{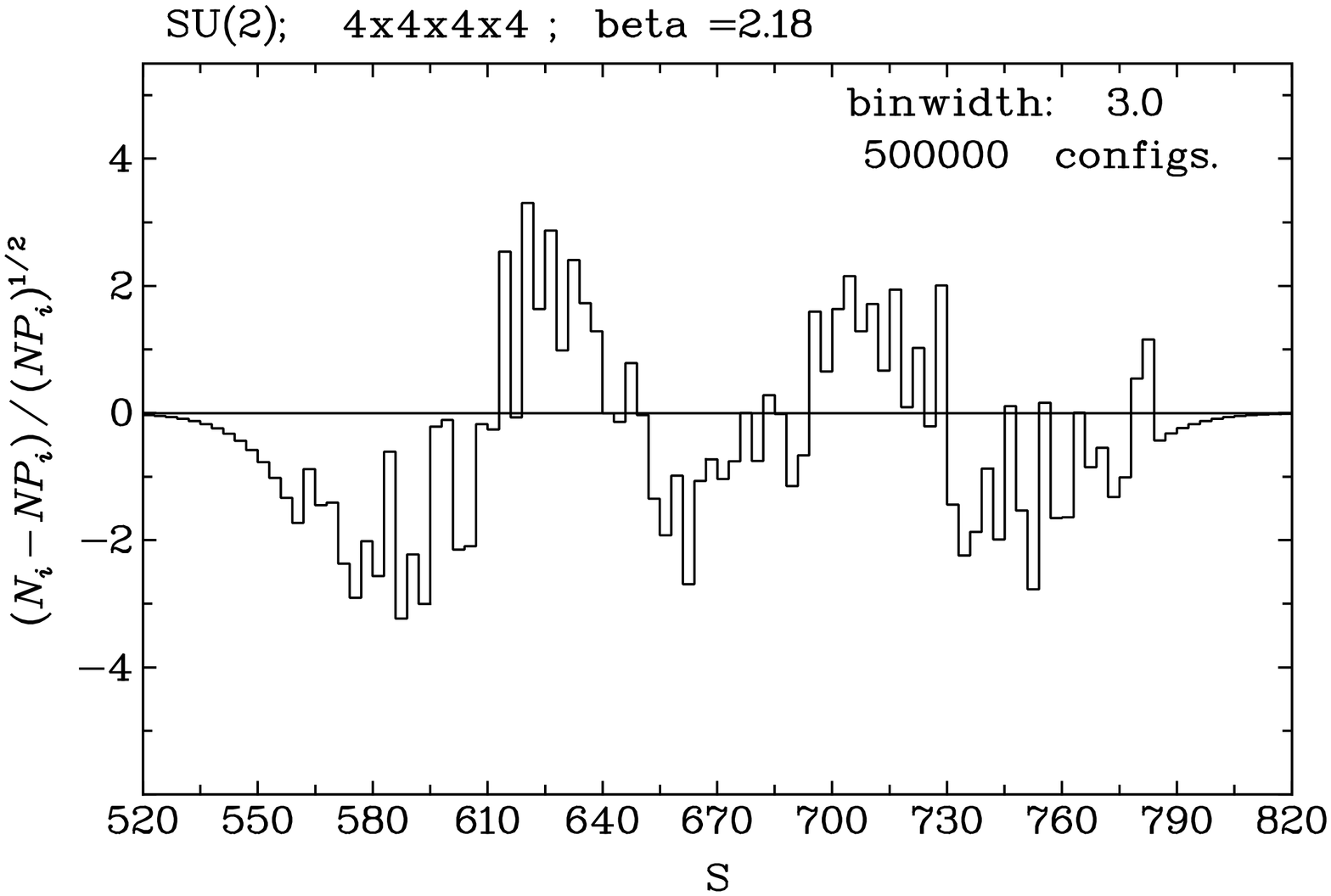}
\includegraphics[width=2.in,angle=270]{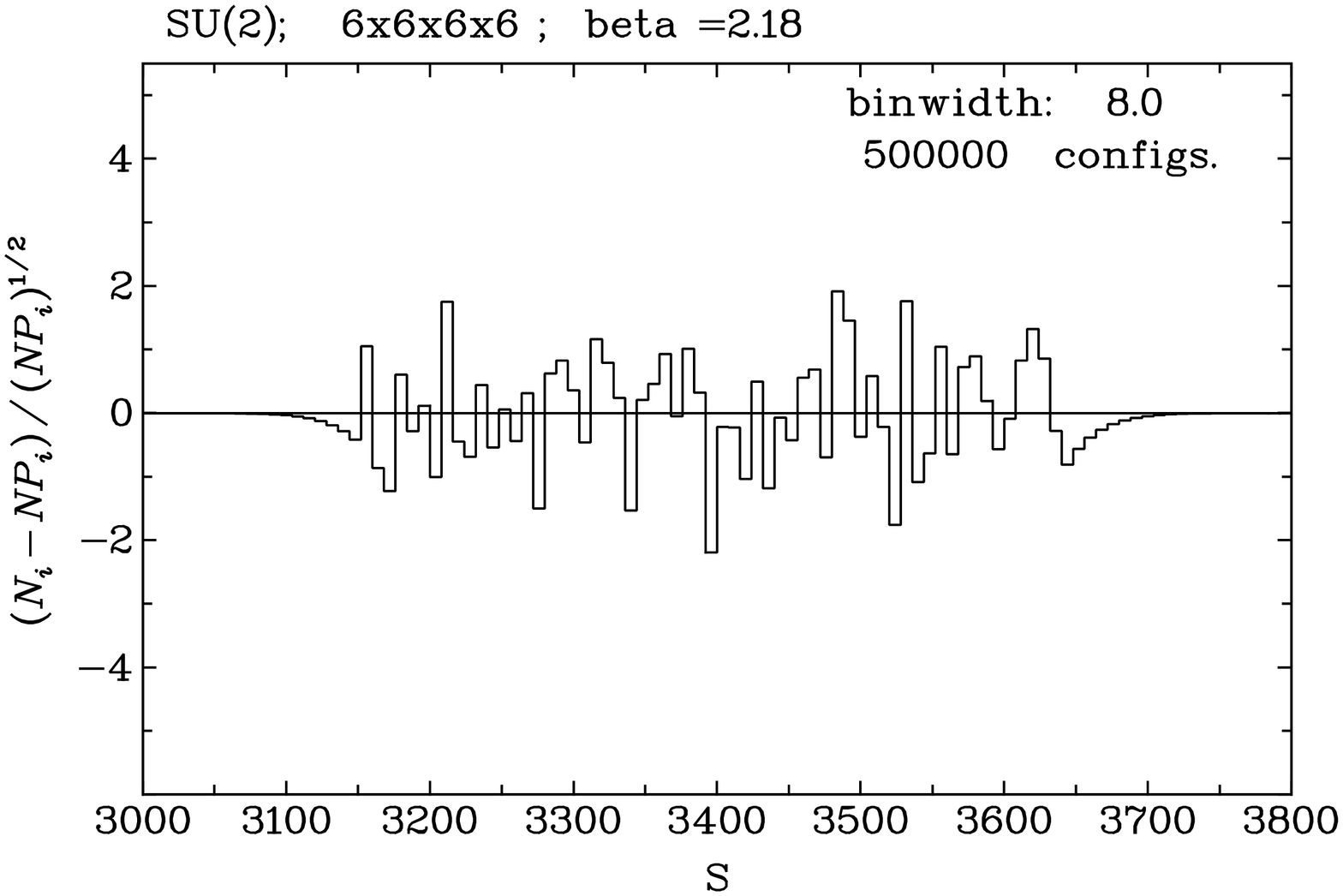}
\vskip20pt
\caption{The residuals $(N_i-NP_i)/(NP_i)^{1/2}$ discussed in the text  for a distribution of 500,000 values of $S$ in an histogram with 100 bins 
for a $SU(2)$ pure gauge theory on a $4^4$ lattice at $\beta =2.18$.}
\label{fig:su2hist}
\end{figure}

In the following, we consider symmetric $4^4$ and $6^4$ lattices. For values of $\beta$ 
near 2, 
the distribution of $S$ is nearly Gaussian and the location of the peak  scales with  the number of sites. The departure from a Gaussian distribution is hardly visible on a histogram. However, as shown in Fig. \ref{fig:su2hist}, the residuals show a coherent behavior on a $4^4$ lattice. As the volume increases, the non-Gaussian features are scaled down and 
for $\beta =2.18$ it seems that the signal is lost in the 
statistical noise. In this figure, $N_i$ is the number of data points in the $i$-th bin and $P_i$ the corresponding probability for a Gaussian distribution with the estimated mean and variance. As in the Gaussian approximation there are no complex zeros. It is crucial 
to resolve the departure from this approximation. 
We now discuss two methods, one based on the estimation of the moments and the other on numerical calculation of the the density of states.  
\section{The Moments Method}
\begin{figure}
\vskip-30pt
\begin{center}
\includegraphics[width=3.5in,angle=270]{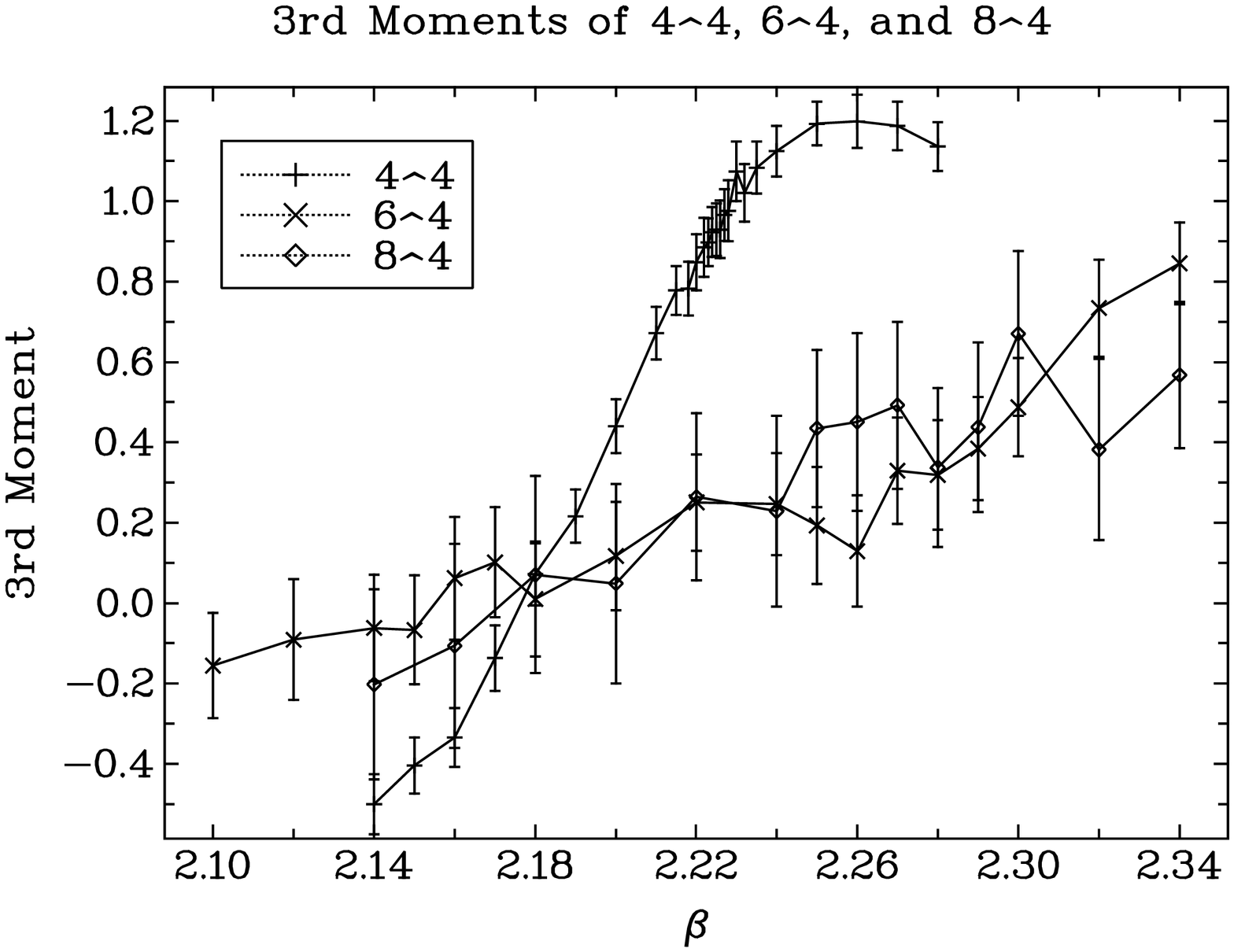}
\vskip-20pt
\includegraphics[width=3.5in,angle=270]{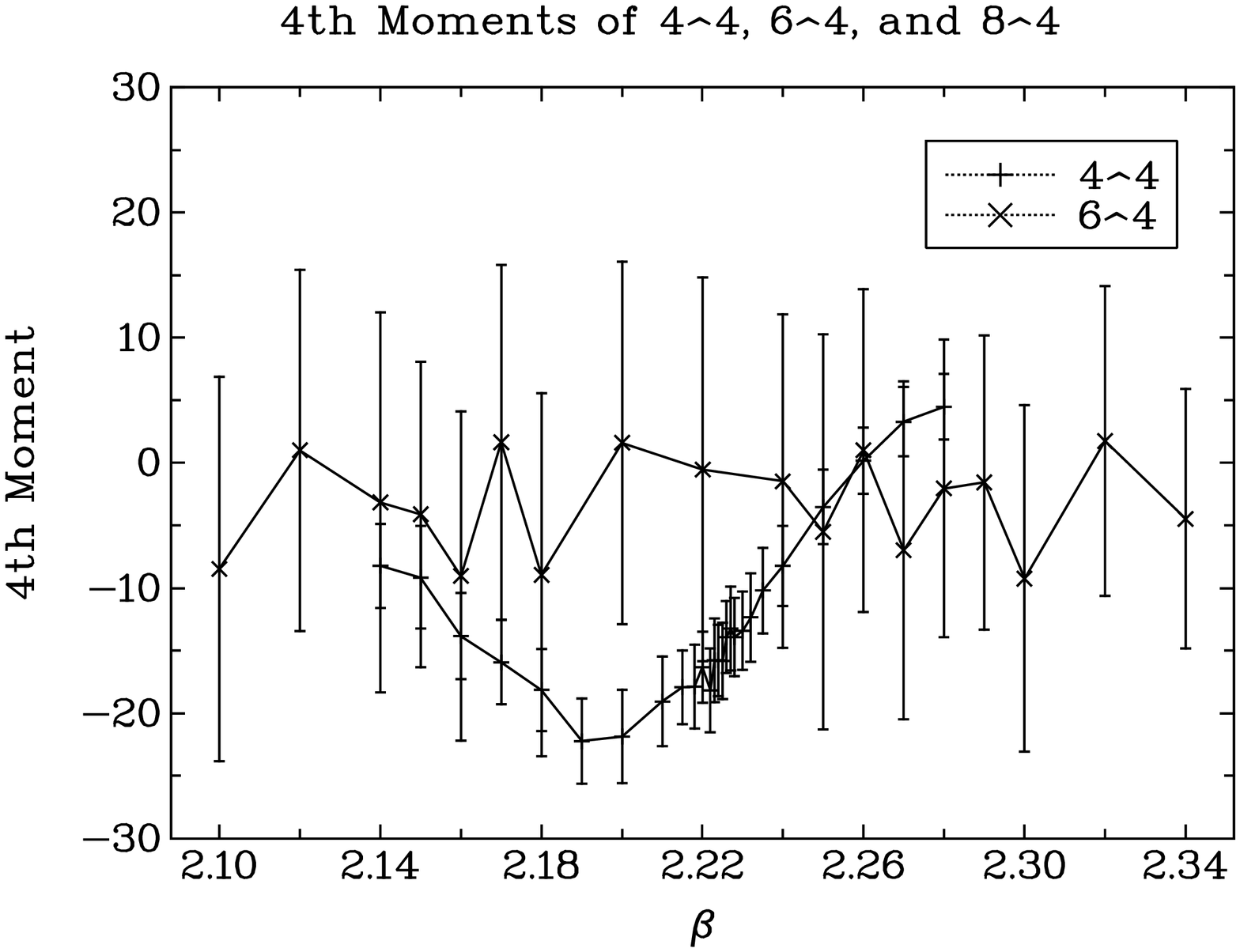}
\end{center}
\caption{$m_3$ and $m_4$ for SU(2) as a function of $\beta$ on $4^4$ and $6^4$ lattices.}
\label{fig:3and4mom}
\end{figure}

\begin{figure}
\centering{\includegraphics[width=3in,angle=270]{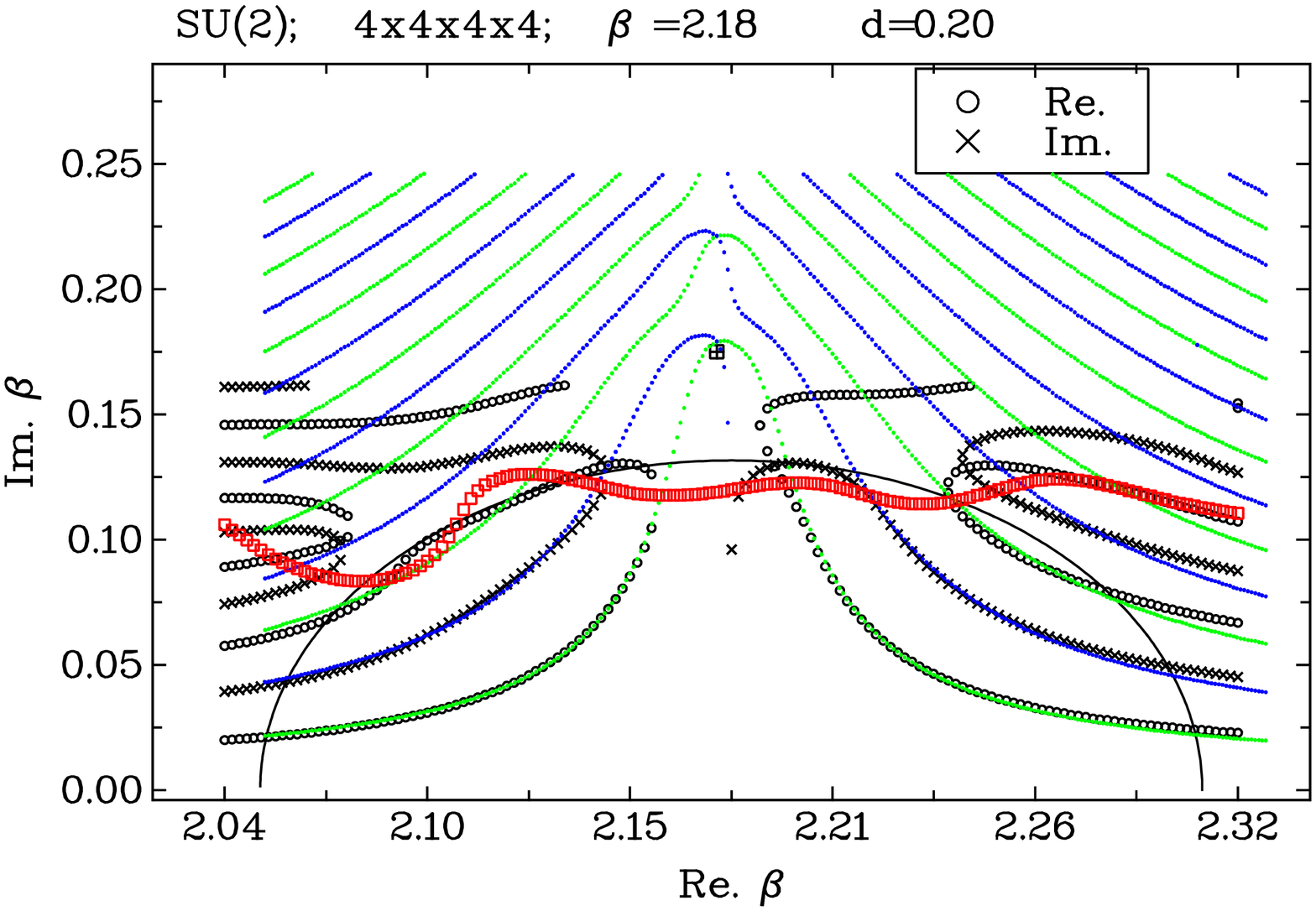}}
\centering{\includegraphics[width=3in,angle=270]{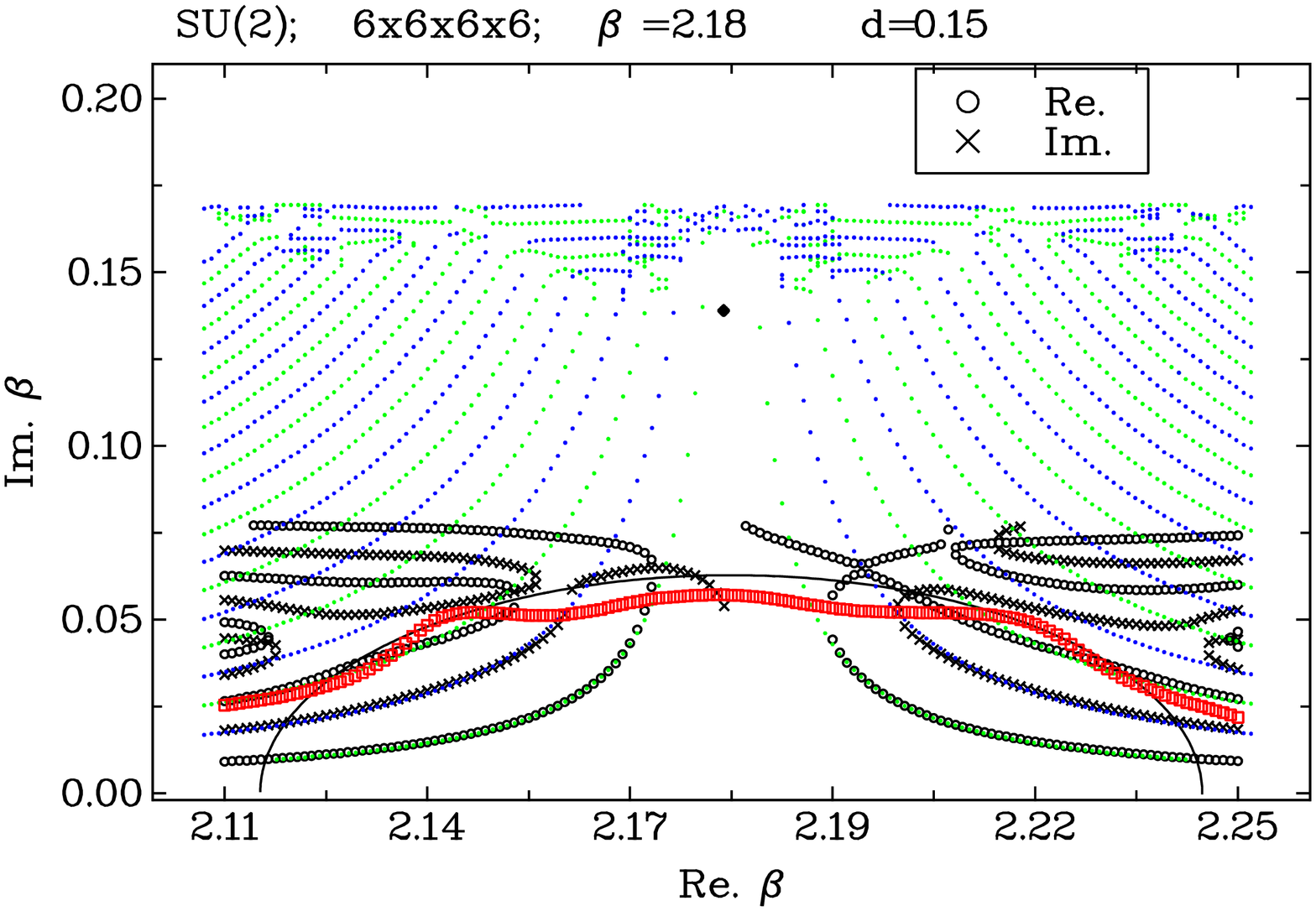}}
\caption{Zeros of the real (crosses) and imaginary (circles) using MC on a $4^4$ lattice,
for $SU(2)$  at $\beta_0 = 2.18$. The smaller dots are the values for the real (green) and imaginary (blue) parts
obtained from the 4 parameter model. The MC exclusion region boundary for $d= 0.20$ (defined in \cite{quasi}) is represented by boxes (red).
}
\label{fig:su}
\end{figure}

\begin{figure}
\includegraphics[width=2.9in,angle=0]{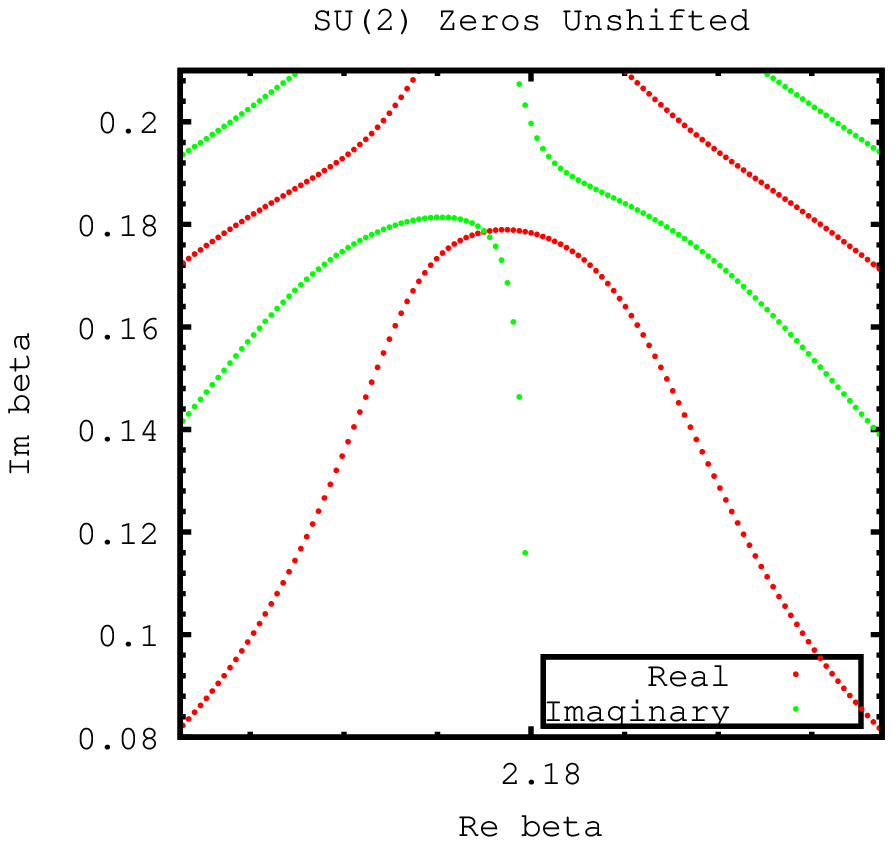}
\includegraphics[width=2.9in,angle=0]{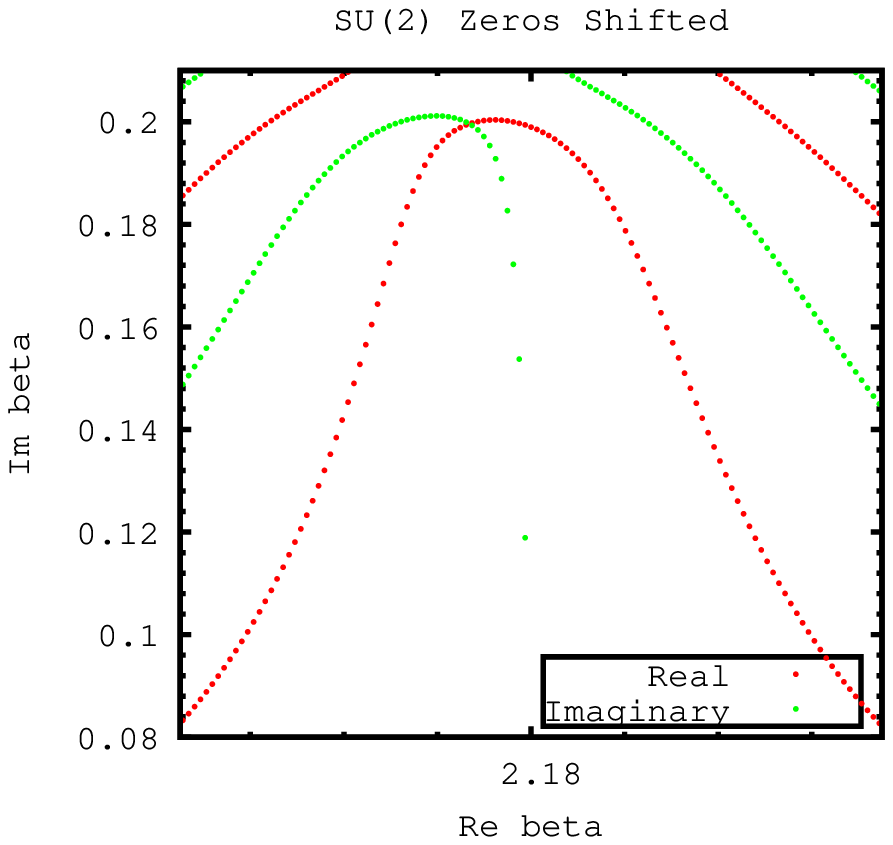}
\caption{Changes of the zeros when the fourth moment is shifted in the positive direction
until the error bar is reached  for $\beta_0=2.18$ on a $4^4$ lattice.}
\label{fig:compare}
\end{figure}

In this section, we consider the following corrections \cite{quasi} to the Gaussian approximation:
\begin{equation}
P(S)\propto \exp(-\lambda_1 S-\lambda_2 S^2-\lambda_3 S^3-\lambda_4 S^4)
\end{equation} 
The four unknown parameters can be determined from the first four moments using Newton's method. The moments are defined as
\begin{eqnarray}
m_1&=&<S>/\mathcal{N}_p\nonumber \\
m_2&=&<(S-<S>)^2>/\mathcal{N}_p\nonumber \\
m_3&=&<(S-<S>)^3> /\mathcal{N}_p\nonumber \\
m_4&=&(<(S-<S>)^4>-3<(S-<S>)^2>^2)/\mathcal{N}_p\nonumber \\
\end{eqnarray}
As $<S>$ scales like $\mathcal{N}_p$ and the individual terms of the $n$-th moment
like  $\mathcal{N}_p^n$, each subtraction implies a loss of significant digits which increases with the volume. As shown in Fig. \ref{fig:3and4mom}
the third and fourth moments have large errors even on a $4^4$ lattice. 
Once we obtain $Z(\beta)$, we can calculate the zeros of real and imaginary parts of it separately. The cross points are the Fisher's zeros. The result for ${\beta_0}=2.18$ on a $4^4$ and $6^4$ lattice is shown in Fig. \ref{fig:su}.
The errors of the moments affect  $P(S)$ and  the location of the zeros. A change of the fourth moment  within the error bars produces changes in the zeros illustrated in Fig. \ref{fig:compare}. This change gives an idea of the errors associated with the method. 
\section{Density of state Method}
\begin{figure}
\begin{center}
\includegraphics[width=3.in,angle=0]{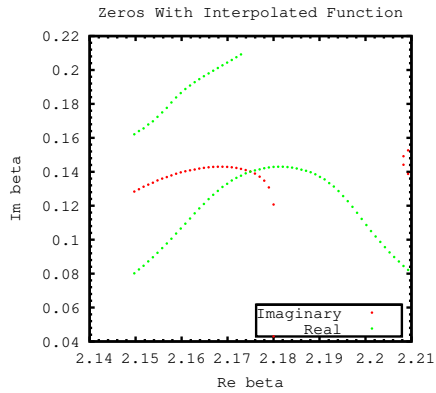}
\end{center}
\caption{\label{fig:inter}Same quantities as in Fig. 4
on a $4^4$ lattice but with 
an interpolated version of $f$.}
\end{figure}

The probability distribution of the plaquette can be written as
\begin{equation}
P_\beta(S)= n(S)\exp(-\beta S)\ .
\end{equation}
To find the $\beta$ independent density of state using Monte Carlo data, we needed to patch the data from different $\beta$  together. First the $\beta$ dependence was removed by multiplying by $e^{\beta S}$.  Using only the bins with statistics higher than half the maximum, we overlay the data from each set on top of one another to make a smooth curve (we took the log of the values in the bins and adjusted the offset with a one-parameter fit).  This procedure can be found with more detail in \cite{dos}. 
Using numerical interpolation for $f(S/\mathcal{N}_p)\equiv ln(n( S/\mathcal{N}_p))/\mathcal{N}_p$, it is possible to calculate the zeros using numerical integration. 
The results are shown in Fig. \ref{fig:inter}. 

As the changes were more important than expected (compared to Fig. \ref{fig:compare}), 
we estimated the errors by using different distributions of $S$.
Three different methods are used to get the partition function, with which $<{\rm cos}(Im\beta(S-<S>))>$ is calculated and compared as a function of the imaginary part of $\beta$ 
at fixed real part 2.18.  
We estimated the error by generating multiple data. For the first, we generated 50 bootstraped sets of configurations and computed $< {\rm cos} ( Im\beta(S-<S>))>$ directly by MC 
average. For the second, we get the partition functions via the density of states which are obtained using interpolation out of 50 patchings. For the last, we fit the 50 patchings using Chebyshev Polynomials instead of interpolation. The Chebyshev fitting seems to have much higher accuracy and stability than the other two methods and will be used 
for further investigations. Calculations of the zeros  for $Im\beta <0.11$ with this 
method appear to be consistent with Fig. \ref{fig:su}.
\begin{figure}
\includegraphics[width=3.in,angle=0]{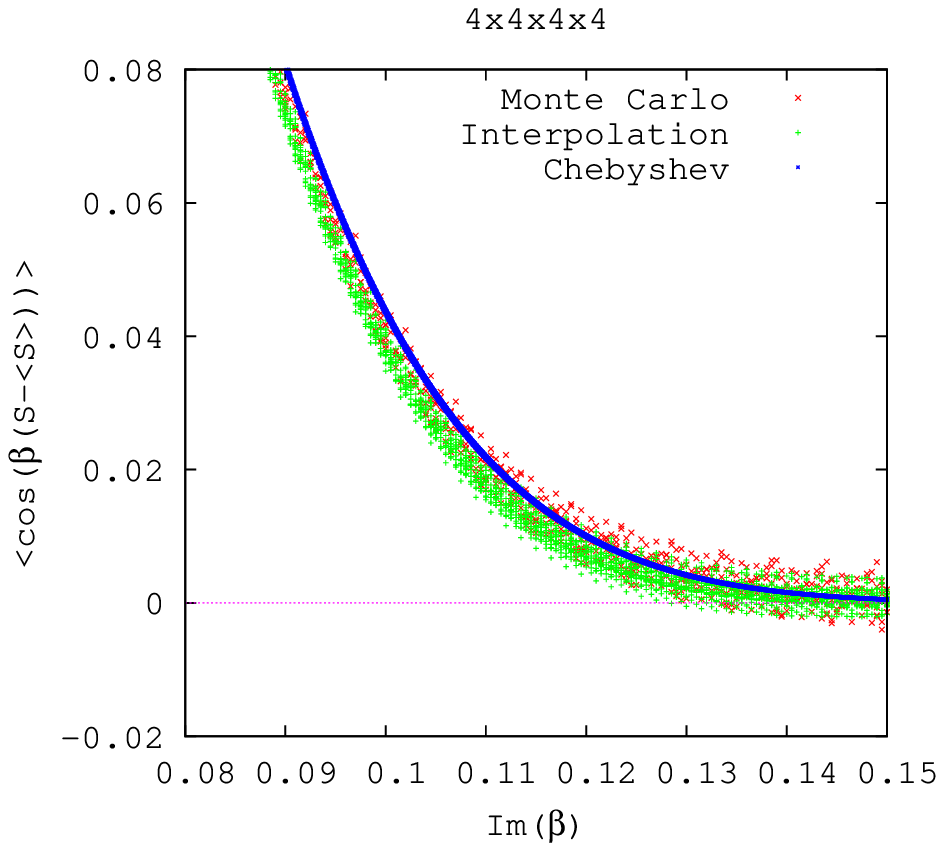}
\includegraphics[width=3.in,angle=0]{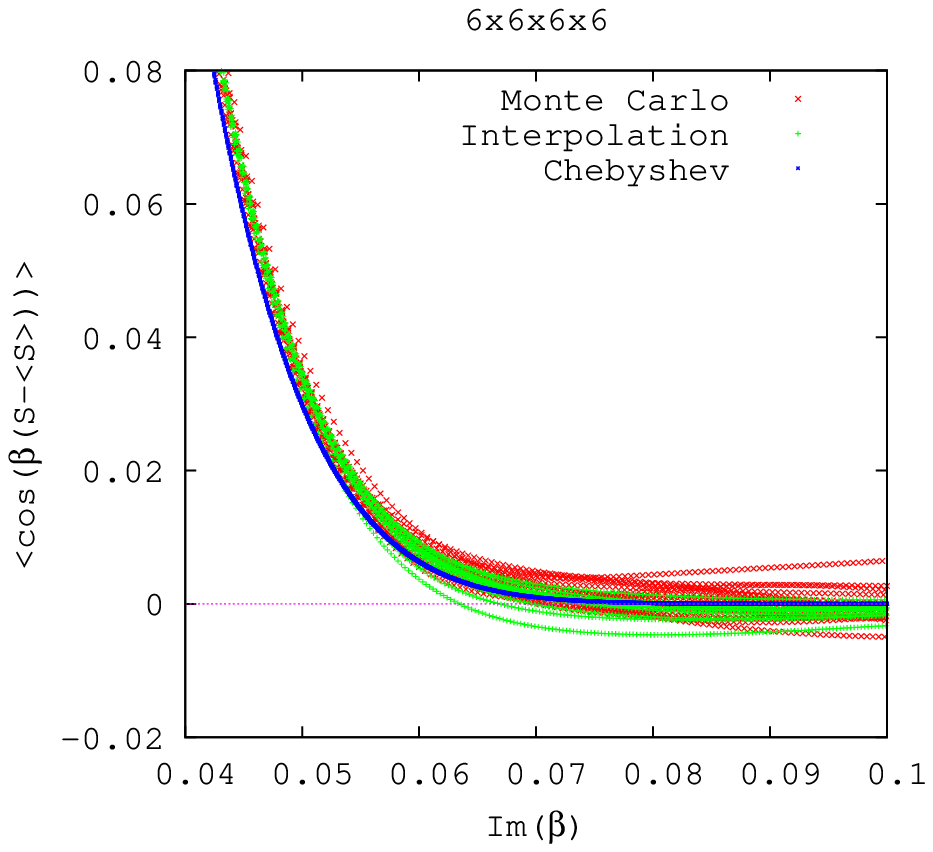}
\caption{ $<{\rm cos}(Im\beta(S-<S>))>$  as a function of the imaginary part of $\beta$ 
at fixed real part 2.18 with
three methods: interpolation, Chebyshev fitting and bootstraped Monte Carlo,  are  on  $4^4$ (left) and $6^4$ (right) lattices . }
\label{fig:cos}
\end{figure}

\section{Conclusions}

In conclusion, we have compared the moments methods and methods based on the 
density of states with the simple MC reweighting procedure to calculate the zeros of 
the partition function. The method where the density of states is approximated by 
Chebyshev polynomials seems the most reliable and will be used in future investigations.

\begin{acknowledgments}
This 
research was supported in part  by the Department of Energy
under Contract No. FG02-91ER40664. 
A.V. work was supported by the Joint Theory Institute funded together by 
Argonne National Laboratory and the University of Chicago, and in part by the U.S. Department of Energy, Division of High Energy Physics and Office of Nuclear Physics, under Contract DE-AC02-06CH11357.

\end{acknowledgments}

\end{document}